\documentclass[12pt,leqno]{article}
\usepackage{amssymb,amsmath,amscd,amsthm,enumerate,epsfig}
\renewcommand\a{\alpha} 
\newcommand\bb{\beta}

\newcommand\f{\phi}
\newcommand\h{\eta} 

\newcommand\la{\lambda}
\newcommand\w{\omega}
\newcommand\dl{\delta}
\newcommand\e{\varepsilon}
\newcommand\Om{\Omega}
\newcommand\G{\Gamma}
\newcommand\g{\gamma}
\newcommand\F{\Phi}

\newcommand\cA{\mathcal{A}}
\newcommand\cB{\mathcal{B}}

\newcommand\cF{\mathcal{F}}
\newcommand\cM{\mathcal{M}}
\newcommand\cN{\mathcal{N}}
\newcommand\cO{\mathcal{O}}

\newcommand\cV{\mathcal{V}}

\newcommand\pb[2]{\left\{#1,#2\right\}} 
\newcommand\pr[1]{\left(#1\right)}        
\newcommand\fpr[1]{\left\{#1\right\}}       

\newcommand\re[1]{\Re\left[{#1}\right]}
\newcommand\im[1]{\Im\left[{#1}\right]}

\newcommand\C{\mathbb C}
\newcommand\R{\mathbb R}
\newcommand\Z{\mathbb Z}

\newcommand\pd{\partial}

\DeclareMathOperator{\Corr}{Corr}

\newcommand\res{\operatorname{res}}
\newcommand\ress{\res_{P_1}}

\newcommand\rr{\mathfrak r} 
\newcommand\ii{\mathfrak i} 
\newcommand\kk{\mathfrak k}
\newcommand\I{\sqrt{-1}}

\newcommand\Tr{T^{\rr}}
\newcommand\Ti{T^{\ii}}
\newcommand\tr{t^{\rr}}
\newcommand\ti{t^{\ii}}
\newcommand\Ta[2]{T^a_{#1,#2}}
\newcommand\Tb[2]{T^b_{#1,#2}}

\newcommand\Ghat{\Hat{\Gamma}_{g}}
\newcommand\Gcut{{\big(\G_{g}\big)}_{\text{cut}}}
\newcommand\GHcut{{\big(\Ghat\big)}_{\text{cut}}}

\newcommand\pal{P_{a_l}}
\newcommand\palp{\pal^+}
\newcommand\palm{\pal^-}
\newcommand\palpm{\palp-\palm}

\newcommand\pbl{P_{b_l}}
\newcommand\pblp{\pbl^+}
\newcommand\pblm{\pbl^-}
\newcommand\pblpm{\pblp-\pblm}

\newcommand\ia[2]{\oint_{a_{#1}}#2}
\newcommand\iap[2]{\oint_{a^{+}_{#1}}#2}
\newcommand\iam[2]{\oint_{a^{-}_{#1}}#2}
\newcommand\iapm[2]{\oint_{a^{\pm}_{#1}}#2}
\newcommand\iaa[2]{\oint_{[a_{#1}]}#2}

\newcommand\iaam[2]{\oint_{[a^{-}_{#1}]}#2}
\newcommand\iaapm[2]{\oint_{[a^{\pm}_{#1}]}#2}
\newcommand\ib[2]{\oint_{b_{#1}}#2}
\newcommand\ibp[2]{\oint_{b^{+}_{#1}}#2}

\newcommand\ibpm[2]{\oint_{b^{\pm}_{#1}}#2}
\newcommand\ibb[2]{\oint_{[b_{#1}]}#2}
\newcommand\ibbp[2]{\oint_{[b^{+}_{#1}]}#2}

\newcommand\ibbpm[2]{\oint_{[b^{\pm}_{#1}]}#2}

\newcommand\wS{(\w_{S})}

\newcommand\oR[1]{\Om^{\rr}_{#1}}
\newcommand\oI[1]{\Om^{\ii}_{#1}}

\newcommand\oah[1]{\Om^a_{h,#1}}
\newcommand\obh[1]{\Om^b_{h,#1}}
\newcommand\oaE[1]{\Om^a_{E,#1}}
\newcommand\obE[1]{\Om^b_{E,#1}}

\newcommand\bm[1]{\begin{bmatrix} #1 \end{bmatrix}}

\DeclareMathOperator{\Log}{Log}

\newtheoremstyle{thm-break}
  {}
  {0pt}
  {\itshape}
  {}
  {\bfseries}
  {.}
  {\newline}
  {}

  \newtheoremstyle{dfn-break}
  {}
  {0pt}
  {}
  {}
  {\bfseries}
  {.}
  {\newline}
  {}

\newtheoremstyle{rem-break}
  {}
  {}
  {}
  {}
  {\itshape}
  {.}
  {\newline}
  {}

\theoremstyle{plain}
\newtheorem{theorem}{Theorem}
\newtheorem*{theorem*}{Theorem}
\newtheorem{proposition}{Proposition}
\newtheorem*{proposition*}{Proposition}
\newtheorem*{corollary}{Corollary} 
\theoremstyle{definition}

\newtheorem*{notation*}{Notation} 
\theoremstyle{thm-break}

\newtheorem*{btheorem*}{Theorem}
\newtheorem{bproposition}[proposition]{Proposition}
\newtheorem*{bproposition*}{Proposition} 
\theoremstyle{dfn-break}

\theoremstyle{remark}
\newtheorem*{rem}{Remark}
\theoremstyle{rem-break}

\begin{document}

\thispagestyle{empty}    
\title{%
Real-normalized Whitham hierarchies\\
and  the WDVV equations}
\author{Anton Dzhamay\\
Department of Mathematics\\
Columbia University\\ 
New York, NY 10027\\
email: ad@math.columbia.edu}
\date{\today}
\maketitle\thispagestyle{empty}

\begin{abstract}
In this paper we present a construction of a new class of explicit solutions to the 
WDVV (or \emph{associativity}) equations. Our construction is based on 
a relationship between the WDVV equations and \emph{Whitham} 
(or \emph{modulation}) equations. 
Whitham equations appear in the perturbation theory of exact 
algebro-geometric solutions of soliton equations and are defined on the 
moduli space of algebraic curves with some extra algebro-geometric data. 
It was first observed by Krichever that for curves of genus zero 
the $\tau$-function of a ``universal'' Whitham hierarchy gives a solution 
to the WDVV equations. This construction was later extended by 
Dubrovin and Krichever to algebraic curves of higher genus.  
Such extension depends on the choice of a normalization for the
corresponding Whitham differentials. Traditionally only 
\emph{complex} normalization (or the normalization w.r.t. 
$a$-cycles) was considered. In this paper we generalize the above 
construction to the \emph{real-normalized} case. 
\end{abstract}

\subsection*{Introduction}\label{S:intro}
In the beginning of the 90's, while studying deformations of $2D$
topological field theories (TFT), E.~Witten, R.~Dijkgraaf,
E.~Verlinde, and H.~Verlinde
(\cite{DVV-topstr,DVV-2Dqgrav,W-topphase}) wrote down the following
overdetermined system of non-linear PDEs for a function
$\cF(t)=\cF(t_{0},t_{1},\dots)$: \begin{equation*}\tag{WDVV}
\cF_{\a\bb\la}(\cF_{0\la\mu})^{-1}\cF_{\mu\g\dl}=
\cF_{\dl\bb\la}(\cF_{0\la\mu})^{-1}\cF_{\mu\g\a},
\end{equation*} where $\cF_{\a}=\dfrac{\pd \cF}{\pd
t_{\a}}$, and the matrix $\h_{\a\bb}=\cF_{0\a\bb}$ is constant and
non-degenerate.  These equations are now called the WDVV equations.
They appear in the following way \cite{Dijk-lesH}.

One can show that the structure of a $2D$ TFT is equivalent to a
structure of a \emph{Frobenius algebra} $A$, i.~e., a commutative
associative algebra with a unit and a symmetric non-degenerate
bilinear form $\h$ such that $\h(a*b,d)=\h(a,b*d)$.  Let $\{\f_{\a}\}$
be a basis of $A$ ($\f_{\a}$ correspond to primary fields in TFT) and
let
$\h_{\a\bb}=\h(\f_{\a},\f_{\bb})=\left<\f_{\a}\f_{\bb}\right>_{0}$,
$c_{\a\bb\g}=\h(\f_{\a},\f_{\bb}*\f_{\g})=\left<\f_{\a}\f_{\bb}\f_{\g}\right>_{0}$.
 Deformations of a TFT structure considered in
\cite{DVV-topstr,DVV-2Dqgrav,W-topphase} correspond to potential
deformations of a Frobenius algebra, i.~e., there should exist a
function $\cF(t)$, called the \emph{WDVV potential}, such that
$\h_{\a\bb}=\pd_{0\a\bb}\cF(t)$ and
$c_{\a\bb\g}(t)=\pd_{\a\bb\g}\cF(t)$.  Then the WDVV equations are
just the associativity conditions for the deformed algebra structure. 

It is now clear that the theory of the WDVV equations
and its differential-geometric counterpart, the theory of Frobenius
manifolds, are related to a whole spectrum of applications ranging from
classical differential geometry (Darboux-Egoroff metrics,
$n$-orthogonal curvilinear coordinate systems, deformations of
singularities) to quantum cohomology, Gromov-Witten invariants,
integrable hierarchies, and the Seiberg-Witten equations.  

Solutions of the WDVV equations that can be obtained from the theory 
of Whitham hierarchies correspond to the topological 
Landau-Ginzburg theories and minimal models. 
For $A_{n}$ minimal models, a Frobenius algebra
structure is defined on the space of degree $n-2$ polynomials in $p$
with the help of the superpotential $W(p)=\dfrac{p^{n}}{n}$
by \begin{align*} u*v &= u(p)v(p) \mod W'(p), \\ \left<uv\right>
&=-\res_{\infty}\frac{u(p)v(p)}{W'(p)} dp.  \end{align*} A Frobenius
algebra structure is then deformed by deforming the superpotential
$W$,
\begin{equation*}
    W(p|a)=\frac{p^{n}}{n}+a_{n-2} p^{n-2}+\cdots+ a_{1}p+a_0,
\end{equation*}
where we used the notation $W(p|a)$ to separate the variable $p$ 
from the deformation parameters $a_{n-2}, \ldots, a_{0}$. 
Let $\kk(p)$ be  an $n^{\text{th}}$ root of $W(p|a)$ 
in the neighborhood of infinity,
\begin{equation*}
\kk^{n}(p)= n W(p)=p^{n}+\cdots, \qquad \kk(p)=p+O(p^{-1}).
\end{equation*}
Then the deformed basis of primary fields $\f_{\a}$ is given by
\begin{equation}
\f_{\a}(p|t)=\frac1{\a+1}\frac{d\Om_{\a+1}(p|t)}{dp}=p^{\a}+\cdots,
\end{equation}
where $\Om_{\a}(p)=\bm{\kk^{\a}(p)}_{+}$ is a principal (i.e., 
polynomial) part of the Laurent expansion of $\kk^{\a}$ in the 
neighborhood of infinity. The flat coordinates $t_{\a}$ on the space 
of the deformation parameters are given by 
$t_{\a}=-\res_{\infty} \kk^{-(\a+1)} p \,dW$, and 
$\f_{\a}(p|t)=-\pd_{t_{\a}}W(p|t)$.
Then
\begin{align*}
\h_{\a\bb}&=\left<\f_{\a},\f_{\bb}\right>
=-\res_{\infty}\frac{\f_{a}\f_{\bb}}{W'(p)}dp=\dl_{\a+\bb,n-2},\\
c_{\a\bb\g}(t)&=(\text{coeff})\sum\limits_{q_{s}|dW(q_{s})=0}\res_{q_{s}}
\frac{d\Om_{\a+1}d\Om_{\bb+1}d\Om_{\g+1}}{dp\,dW}.
\end{align*}
In \cite{DVV-topstr} it was shown that $c_{\a\bb\g}(t)$ satisfy the 
integrability conditions and therefore
$c_{\a\bb\g}(t)=\pd_{\a\bb\g}\cF(t)$.
The closed expression for this WDVV potential
$\cF(t)$ was identified by I.~Krichever \cite{Kr-dispLax} with the
logarithm of the $\tau$-function of a certain reduction of the genus
zero Whitham hierarchy.  

Whitham equations (or modulation equations)
appear in the theory of perturbations of exact algebro-geometric
solutions of soliton equations and describe ``slow drift'' on the
moduli space of algebro-geometric data.  These equations can be
defined with the help of certain differentials $d\Om_{i}$ on the
universal curve, each differential generating a corresponding Whitham
flow on the moduli.  A large class of solutions of the Whitham
hierarchy is given by the so-called
algebraic orbits. 
Such solutions depend only on finitely many parameters and 
can be constructed as follows.  One starts with a
finite-dimensional moduli space, called the universal configuration
space, that consists of a curve $\G$, punctures $P_{\a}$, and a pair
of Abelian integrals $E$ and $Q$.  Then one defines special
coordinates on this space (Whitham times), picks a leaf defined by
fixing some of them, and maps this leaf to the moduli space of
algebro-geometric data in such a way that coordinate lines go to
Whitham flows.  All information about such algebraic orbit can be
encoded in a single function $\tau(t)$ that depends only on the moduli. 
This function is called the $\tau$-function of an algebraic orbit.  

In the genus zero case Abelian integrals become polynomials, and if we
choose an algebraic orbit with $Q=p$, then $E$ can be identified (up
to normalization) with the superpotential $W$, $\Om_{i}$ define a
basis of the corresponding Frobenius algebra, Whitham times $t_{i}$
give flat coordinates on the orbit, and $F(t)=\ln \tau(t)$ is a WDVV
potential.  

This approach can be generalized to moduli spaces of
curves of higher genus.  One new feature of a higher genus case is a
more complicated topology of the moduli space.  As a result, the
hierarchy has to be extended to include certain (multivalued)
differentials $d\Om_{A}$ that generate additional Whitham flows. 
Another important issue is a choice of the normalization. 
Namely, in the genus zero case, differentials $d\Om_{i}$ are completely
determined by their expansions in the neighborhoods of marked points
$P_{\a}$.  In the higher genus case, these conditions define $d\Om_{i}$
only up to a holomorphic differential.  This ambiguity is fixed by
introducing a normalization condition.  There are two main choices ---
real normalization, which is defined by 
\begin{equation*}
\im{\oint_{c}d\Om_{i}}=0\qquad \forall c\in H_{1}(\G_{g},\Z),
\end{equation*} 
and complex normalization (or normalization w.r.t.
$a$-cycles).  Complex normalization requires making a choice of a
canonical basis $\cB$ of cycles in the homology of $\G_{g}$ and is
then defined by 
\begin{equation*} \oint_{a_{k}}d\Om_{i}=0\qquad
\forall a_{k}\in\cB. 
\end{equation*} Whitham equations were
originally derived in \cite{Kr-aver} for real-normalized
differentials.  However, after the relationship between Whitham
equations and WDVV equations was found in \cite{Kr-dispLax} and
\cite{Dub-whithm}, the focus shifted to the complex normalization
condition \cite{Kr-tauWh}.  There were two main reasons for this. 
First, in the complex-normalized case the $\tau$-function is
holomorphic, which is important for string theory applications. 
Second, the derivation of the expression for the $\tau$-function
relied on the Riemann bilinear identities.  Corresponding identities
for real normalized differentials are technically more complicated. 
However, complex normalization has certain disadvantages.  In
particular, it is well-defined only on the extended moduli space that
incorporates the choice of a canonical basis $\cB$ into moduli data. 

In this paper we develop a real-normalized version of the above
approach.  In this case, Whitham hierarchy can be defined on the
usual moduli space of curves with some extra algebro-geometric data. 
First, we define a real leaf in the universal configuration space,
introduce Whitham coordinates on this leaf and map it into the moduli
space in such a way that the resulting differentials are
real-normalized.  Then we prove the real-normalized version of the
Riemann bilinear identities (generalized to multivalued
differentials).  Using these identities we find the formula for the
$\tau$-function of an algebraic orbit and prove the following theorem
for $F(T)=\ln \tau(T)$.  
\begin{theorem*} The third derivatives of $F(T)$ are given by the following
formula \begin{equation*} \pd_{T_A T_B T_C} F(T)= \re{\sum_{q_s |
dE(q_s)=0} \res_{q_s}{\frac{d\Om_A d\Om_B d\Om_C}{dE dQ}}}
\end{equation*} 
\end{theorem*}
This theorem then implies that if we consider a special class of
algebraic orbits by choosing $dQ$ to be a
real-normalized differential with a pole of order two at a puncture
$P_{1}$, then Whitham flows define potential deformations of a
Frobenius algebra structure with observables corresponding to
$\dfrac{d\Om_{A}}{dQ}$, and we have the following theorem.
\begin{theorem*}
    The logarithm of a $\tau$-function of a (reduced) 
    real-normalized genus $g$ Whitham hierarchy
    is a solution to the WDVV equation.
\end{theorem*}

\paragraph*{Acknowledgments} 
The author is grateful to I.~Krichever for suggesting this problem and 
for constant attention to this work. The author also thanks 
Yu.~Volvovski for many useful discussions.

\subsection*{Whitham equations}\label{ss:We}

Whitham equations first appeared in the theory of perturbations of exact algebro-geometric 
solutions of soliton equations. Such solutions are defined by linear 
flows on the Jacobian $\operatorname{Jac}(\G_{g})$ of an auxiliary 
algebraic curve $\G_{g}$ and are expressed in terms of theta functions. 
Perturbing these solutions by the so-called non-linear WKB 
(or Whitham averaging~\cite{Wh}) method 
(\cite{FFM,DobrMasl-multphas,DubNov-BogWh}) results in a ``slow drift'' 
on the moduli $\cM$ of algebro-geometric data. Equations describing 
this drift are called \emph{Whitham equations}.

Although Whitham equations are equations on the moduli, 
they can be conveniently written with the help of
certain Abelian differentials $d\Om_{A}(P|I)$, $P\in\G_{g}$ defined on 
the universal curve $\cN_{g}^{1}$,
\begin{equation*}\begin{CD}
\G_{g} @>>>  \cN_{g}^{1}\\
@. @VVV\\
@. \cM.
\end{CD}\end{equation*}

Each of the differentials $d\Om_{A}$ is coupled with a corresponding 
\emph{Whitham time $T_{A}$} defining
$A^{\text{th}}$ Whitham flow on $\cM$. Then, after making a special choice of 
connection on $\cN_{g}^{1}$,
Whitham equations can be written in the following implicit form,
\begin{equation*}
\pd_{A}d\Om_{B}=\pd_{B}d\Om_{A}.\label{eq:wh}
\end{equation*}
This form of Whitham equations was first observed by Flashka, Forest, and
McLaughlin \cite{FFM} for the KdV equation and hyperelliptic
spectral curves. It was later justified by Krichever \cite{Kr-aver} in a more
general setting of $(2+1)$ equations and general spectral curves. 

\subsection*{Whitham hierarchies}\label{ss:Wh}
It turns out that one can construct a whole hierarchy of 
Whitham equations. Following Krichever, we use 
algebro-geometric approach to construct Whitham hierarchies 
(Hamiltonian approach to the theory of 
Whitham equations was developed in~\cite{DubNov-BogWh}).

We begin with a local definition. 
Let $T=\{T_{A}\}_{A\in\cA}$ be a collection of (real or complex) times, 
indexed by some set $\cA$, let $z\in D\subset\C$, and let $\{\Om_{A}(z|T)\}_{A\in\cA}$ 
be a collection of functions, 
meromorphic in $z$, each $\Om_{A}$ is coupled with the corresponding time $T_{A}$. 
The functions $\Om_{A}(z|T)$
should be thought of as pull-backs via Whitham times of Abelian integrals 
$\Om_{A}=\int d\Om_{A}$ from 
a leaf spanned by Whitham flows in $\cM$, 
with $z$ corresponding to a local coordinate along the fiber. 
Define a $1$-form $\w$ by
\begin{equation*}
\w=\sum_{A} \Om_{A}(z|T)\,dT_{A}.
\end{equation*}
Then, by definition, \emph{Whitham hierarchy} is given by the generating equation
\begin{equation}
\dl\w\wedge\dl\w=0,\label{eq:wh-hier}
\end{equation}
where
\begin{equation*}
\dl\w=\sum_{A} \pd_{B} \Om_{A}\,dT_{B}\wedge dT_{A} +\frac{d\Om_{A}}{dz}\,dz\wedge dT_{A}.
\end{equation*}
This is equivalent to the following two equations,
\begin{equation}
\sum_{\{A,B,C\}}\e^{\{A,B,C\}}\left(\frac{d\Om_A}{dz}\right)(\pd_B\Om_C)=0,\qquad 
\sum_{\{A,B,C,D\}}\e^{\{A,B,C,D\}}(\pd_{A}\Om_{B})(\pd_{C}\Om_{D})=0\label{eq:dag},
\end{equation}
where we sum over all possible permutations of a fixed collection of indexes and 
$\e$ is a sign of a permutation.

Usually we have one marked index $A_{0}\in\cA$ with 
\begin{equation*}
X=T_{A_{0}},\qquad p(z|T)=\Om_{A_{0}}(z|T).
\end{equation*}
Then, as long as $\dfrac{dp}{dz}\neq0$, we can make a change of coordinates 
from $(z,T)$ to $(p,T)$. In these
coordinates, equation \eqref{eq:dag} written for $A_{0}$, $A$, $B$  takes the form
\begin{equation}\label{eq:zc-Wh}
\pd_{A}\Om_{B}(p|T)-\pd_{B}\Om_{A}(p|T)+\pb{\Om_{A}}{\Om_{B}}(p|T)=0,
\end{equation}
where
\begin{equation*}
\pb{\Om_{A}}{\Om_{B}}(p|T)=\pd_{X}\Om_{A}\frac{d\Om_{B}}{dp}-\pd_{X}\Om_{B}\frac{d\Om_{A}}{dp}
\end{equation*}
is the usual Poisson bracket. Equations in this form are called \emph{zero-curvature equations}.
They can be interpreted as a compatibility conditions for the following Hamiltonian system,
\begin{equation*}
\pd_{A} E(p|T)=\pb{E}{\Om_{A}}(p|T).
\end{equation*}
Then, if zero-curvature equations are satisfied, there exists (locally) a solution 
$E=E(p|T)$ of this system 
and, as long as $\dfrac{dE}{dp}\neq0$, we can again change coordinates 
from $(p,T)$ to $(E,T)$. In these coordinates,
the above system takes the form
\begin{equation*}
\pd_{A} p(E|T)=\pd_{X}\Om_{A}(E|T)
\end{equation*}
and its compatibility conditions can be written as
\begin{equation*}
\pd_{A}\Om_{B}(E|T)=\pd_{B}\Om_{A}(E|T).
\end{equation*}
Therefore, there exists (locally) a function $S=S(E|T)$ such that $\Om_{A}(E|T)=\pd_{A}S(E|T)$. 
This function
$S(E|T)$ is called a \emph{prepotential of the Whitham hierarchy}.

\subsection*{Algebraic orbits}\label{ss:algorb}

One can use the above formalism to construct certain exact solutions of 
Whitham equations. These solutions depend on finitely many parameters 
and are obtained as follows \cite{Kr-tauWh}. First we construct
Whitham flows on finite-dimensional 
submanifolds, called \emph{algebraic orbits}, of 
$\cM$ (note that $\cM$ is usually infinite-dimensional) together with Whitham equations 
that they satisfy. Then these
equations are extended to the whole $\cM$. By making a correct choice of algebraic data, 
one can obtain usual Whitham equations of the soliton theory.

We illustrate this idea in the special case of the so-called 
\emph{dispersionless Lax equations}.  
In this case we take algebraic curve of genus zero, $\Gamma\simeq\C P^{1}$, with a 
single puncture $P_{1}=\infty$. The 
moduli space is equivalent to a moduli space of local coordinates 
around $P_{1}$,
\begin{equation*}
\cM=\hat{\cM}_{0,1}=\fpr{\Gamma\simeq\C P^{1};P_{1}=\infty;z(P)}\simeq\fpr{\nu_{s},s=1,\dots},
\end{equation*}
where
\begin{equation*}
z^{-1}(p)=\kk(p)=p+\nu_{1}p^{-1}+\nu_{2}p^{-2}+\cdots,
\end{equation*}
and $p$ is the standard coordinate on $\C$. Note that in this case $\Om_{1}=p$, i.e., 
the marked index is $i=1$.
Then, by definition, the dispersionless Lax hierarchy (or 
dispersionless KP hierarchy) is a set of evolution equations on $\nu_{s}=\nu_{s}(T)$:
\begin{equation*}
\pd_{i}\kk(p|T)=\pb{\kk}{(\kk^{i})_{+}}=\pb{\kk}{\Om_{i}}, 
\qquad\text{where }\pd_{i}=\frac{\pd}{\pd T_{i}}.\label{eq:dKP}
\end{equation*}
This hierarchy can be thought of as a quasi-classical limit of 
the usual KP hierarchy, and algebraic orbits correspond to 
$n^{\text{th}}$ order reductions of the KP hierarchy, i.e., 
nKdV hierarchies.
Namely, let 
\begin{equation*}
E(p)=p^{n}+u_{n-2}p^{n-2}+\cdots+u_{0}
\end{equation*}
and define $\kk(p)$ by the condition $\kk(p)^{n}=E(p)$. 
Then $\nu_{s}=\nu_{s}(u_{0},\dots,u_{n-2})$ and we obtain
a finite-dimensional leaf in $\cM$. 
Corresponding evolution equations on $u_{i}=u_{i}(T)$ can be written as
\begin{alignat*}{2}\label{eq:evolu}
\pd_{i}E(p|T)&=\pb{E}{(E^{\frac in})_{+}}&\qquad&\text{in $(p,T)$ coordinates},\\
\pd_{i}p(E|T)&=\pd_{X}\Om_{i}(E|T)&\qquad &\text{in $(E,T)$ coordinates.}
\end{alignat*}
Then the solution $E=E(p|T)$ (i.e., $u_{i}=u_{i}(T)$) can be obtained as follows 
\cite{Kr-aver,Kr-dispLax}.

\begin{theorem*}\label{thm:g0}
Let 
\begin{equation}
S(p|T)=\sum_{i}T_{i}\Om_{i}(p|u)=\sum_{i}T_{i}\kk^{i}+O(\kk^{-1}).
\end{equation}
Require that 
\begin{equation}\label{eq:ondS}
\frac{dS}{dp}(q_{s})=0\text{ for all }q_{s} \text{ such that }\frac{dE}{dp}(q_{s})=0.
\end{equation}
Equation \eqref{eq:ondS} is equivalent to a collection of equations $F_{k}(u,T)=0$ 
that implicitly define $u_{i}=u_{i}(T)$.

Then $u(T)$ is a solution of the dispersionless Lax equations.
\end{theorem*}

The theorem follows from the fact that $S$ defined above is a \emph{global} prepotential, i.e., 
\begin{equation*}
\pd_{i} S(E|T)=\Om_{i}(E|T)=\begin{cases} \kk^{i} +O(\kk^{-1}) \text{ near } P_{1}\\
\text{has no other poles}\end{cases}.
\end{equation*}

From the definition of $S$ it is clear that the first condition is satisfied. The role of the
condition \eqref{eq:ondS} is to ensure that $\pd_{i} S(E|T)$ is holomorphic on $\G-P_{1}$. 

Note that the condition \eqref{eq:ondS} can be also written in the form $dS=Q dE$ 
for some polynomial $Q(p|T)$. Conversely, 
we can recover $T_{i}$ from $S$ and $\kk$ by the formula
\begin{equation*}
T_{i}=-\frac1i\res_{P_{1}} \kk^{i}\,dS.
\end{equation*}
This observation motivates the following alternative approach. Consider the so-called 
\emph{universal configuration space}
\begin{equation*}
\cM_{0}(n,m)=\fpr{\G=\C P^{1}; P_{1}=\infty; [z]_{n}; E,Q}
\simeq\fpr{u_{0},\dots,u_{n-2};b_{0},\dots,b_{m}},
\end{equation*}
where 
\begin{equation*}
E(p)=p^{n}+u_{n-2}p^{n-2}+\cdots +u_{0},\qquad 
Q(p)=b_{m}p^{m}+\cdots +b_{0},
\end{equation*}
$[z]_{n}$ is an $n$-jet of a local coordinate near $P_{1}$ and we require that
$E=z^{-n} + O(z)$
near $P_{1}$. By definition, put
\begin{equation}\label{eq:dS=QdE}
dS=Q\,dE, \qquad T_{i}=-\frac1i\res_{P_{1}} z^{i}dS.
\end{equation}
Then
\begin{equation*}
\pd_{i} dS(E|T)=d\Om_{i}(E|T),
\end{equation*} 
condition \eqref{eq:dS=QdE} is equivalent to a collection of equations 
$T_{k}=T_{k}(u,b)$,
which can be inverted,
$u_{i}=u_{i}(T)$, $b_{j}=b_{j}(T)$,
and therefore we obtain a map from $\cM_{0,1}$ to an algebraic orbit in $\hat{\cM}_{0}$.

\subsection*{$\tau$-functions and solutions to the WDVV equations}\label{ss:taufn}

For each algebraic orbit constructed above corresponds a so-called $\tau$-function. By definition,
\begin{equation*}
\ln \tau(T)=F(T),
\end{equation*}
where
\begin{equation}
F(T)=\frac12\res_{\infty}\pr{\sum_{i}T_{i}\kk^{i}}\,dS,
\end{equation}
and $dS$ is a (differential of) the prepotential of the corresponding algebraic orbit. 
Then we have the following theorem (Krichever \cite{Kr-dispLax}).

\begin{theorem*}
The derivatives of $F(T)$ are given by the following formulas:
\begin{align*}
\pd_{i} F(T)&=\res_{\infty} \kk^{i}\,dS\\
\pd_{ij} F(T)&=\res_{\infty} \kk^{j}\, d\Om_{i}=\res_{\infty} \kk^{j}\, d\Om_{i}\\
\pd_{ijk} F(T)&= \sum_{q_{s}|dE(q_{s})=0} \res_{q_{s}}\frac{d\Om_{i}d\Om_{j}d\Om_{k}}{dQdE},\qquad 
\text{where }Q=\frac{dS}{dE}.
\end{align*}
\end{theorem*}

\begin{corollary}
If we choose 
\begin{equation*}
Q(p|T)=p,\qquad E(p|T_{1},\dots,T_{n+1})=p^{n}+u_{n-2}(T)p^{n-2}+\cdots+u_{0}(T),
\end{equation*}
then $F(T)$ is a WDVV potential for $A_{n-1}$ Landau-Ginzburg model defined by a superpotential
\begin{equation*}
W(p|t_{0},\dots,t_{n-2})=\frac1n E(p|t_{0},\frac{t_{1}}2,\dots,\frac{t_{n-2}}n,0,\frac1{n+1}).
\end{equation*}
\end{corollary}

\subsection*{Higher genus case}\label{ss:hg-cplx}

The above approach can be generalized to algebraic curves of higher genus.
This question was considered by Dubrovin ~\cite{Dub-whithm}
for Hurwitz spaces and by Krichever~\cite{Kr-tauWh} in general. 
In the higher genus case, the ``universal'' moduli space 
\begin{equation*}
\Hat{\cM}_{g,N}=\fpr{\G_g;P_{\a};\kk^{-1}_{\a}(P)=z_{\a}(P)}
\end{equation*}
consists of the following algebro-geometric data:
\begin{itemize}
\item smooth algebraic curve $\G_{g}$ of genus $g$,
\item collection of marked points $P_{\a}\in\G_g,\quad\a=1,\dots,N$,
\item local coordinates $z_{\a}(P)$ in the neighborhoods 
  of $P_{\a}$, $z_{\a}(P_{\a})=0$.
\end{itemize}
For simplicity, we concentrate exclusively on a single puncture case, $N=1$.

To construct a realization of a Whitham hierarchy on $\Hat{\cM}_{g,1}$, one can
proceed as follows (see~\cite{Kr-tauWh}).
Instead of polynomials $\Om_{i}$ one has to consider 
Abelian differentials of the second kind $d\Om_{i}$ on $\G_{g}$ 
with prescribed behavior near the puncture,
\begin{equation}
d\Om_{i}=d\big(\kk^{i}+O(\kk^{-1})\big)\qquad\text{near $P_{1}$.}\label{eq:dOmi}
\end{equation}
However, condition \eqref{eq:dOmi} specifies $d\Om_{i}$ only up to a holomorphic differential.
To define $d\Om_{i}$ uniquely, one has to impose certain normalization conditions.
There are two choices --- \emph{real normalization} and \emph{complex normalization} 
(or normalization w.r.t. \emph{$a$-cycles}). Real normalization is defined by
the condition
\begin{equation}
\im{\oint_{c}d\Om_{i}}=0\qquad \forall c\in H_{1}(\G_{g},\Z).\label{eq:imnorm}
\end{equation}
To define complex normalization, one has to first choose a canonical homology basis for
 $\G_g$,
\begin{equation*}
    \cB=\fpr{a_{1},\dots,a_{g};b_{1},\dots,b_{g}\ |\ a_{i},b_{j}\in H_{1}(\G_{g},\Z),
    a_{i}\cdot a_{j}=b_{i}\cdot b_{j}=0,a_{i}\cdot b_{j}=\dl_{ij}}.
    \label{eq:cB}
\end{equation*}
As a result, it is necessary to consider the extended moduli space
\begin{equation*}
\Hat{\cM}_{g,1}^*=\fpr{\G_g; P_1; z(P); \cB },
\end{equation*}
which is a covering of $\Hat{\cM}_{g,1}$.
The differentials $d\Om_{i}$ are then uniquely
normalized by the condition
\begin{equation}
\oint_{a_{k}}d\Om_{i}=0\qquad \forall a_{k}\in\cB.
\end{equation}
As before, the marked index is $i=1$ and $p(P)=\int^{P}d\Om_{1}$. 

\subsection*{Universal configuration space}
In  order to construct solutions 
corresponding to algebraic orbits, it is convenient to use the 
universal configuration space approach.
Following \cite{KrPh-soli2susy,KrPh-symplfrms}, define the universal configuration 
space for Whitham hierarchies,
\begin{equation*}
\cM_g(n,m)=\left\{\G_g;P_1;[z]_{n};E,Q\right\}, 
\end{equation*}
to be the moduli space of the following data: 
\begin{itemize}
\item genus $g$ Riemann surface $\G_g$,
\item marked point $P_1\in \G_{g}$,
\item an $n$-jet $[z]_{n}$ of local coordinates $z(P)$ in the neighborhoods 
  of $P_{1}$, $z(P_{1})=0$,
\item an Abelian integral $E$ with pole of order $n$ at $P_{1}$ such that
\begin{equation}
    dE\sim d(z^{-n}+O(z))\qquad \text{near $P_{1}$,\qquad $z\in[z]_{n}$ },
    \label{eq:dE and z}
\end{equation}
\item an Abelian integral $Q$ with pole of order $m$ at $P_{1}$.
\end{itemize}
More precisely, by Abelian integrals we mean pairs
    $E=(dE,P_{0}^{E})$, $Q=(dQ,P_{0}^{Q})$,
where $dE$, $dQ$ are meromorphic differentials of the second kind on $\G_{g}$,
holomorphic on $\G_{g}-P_{1}$, and with poles of order $n+1$ and $m+1$ at $P_{1}$, and
\begin{equation*}
    E(P)=\int_{P_{0}^{E}}^{P}dE,\qquad Q(P)=\int_{P_{0}^{Q}}^{P}dQ.
    \label{eq:EQint}
\end{equation*}
 Alternatively, one can choose a local coordinate
$z\in [z]_{n}$ and define $E$ and $Q$ to be pairs $E=(dE,c_{E})$, 
$Q=(dQ,c_{Q})$, where near $P_{1}$,
\begin{equation*}
    E\sim z^{-n}+c_{E}+O(z),\qquad 
    Q\sim c_{m}^{Q}z^{-m}+\cdots+c_{1}^{Q}z^{-1}+c_{Q}+O(z).
\end{equation*}
Note that after $E$ is chosen, there is a preferred local coordinate
$z_{*}\in[z]_{n}$ defined by
\begin{equation*}
    E= z_{*}^{-n}\qquad \text{near $P_{1}$}.
    \label{eq:z*}
\end{equation*}
The moduli space $\cM_{g}(n,m)$ is a complex manifold of dimension 
$5g+n+m$ with at most orbifold singularities.
We also need to consider smaller $(4g+n-1)$-dimensional moduli space
\begin{equation*}
    \cM_{g}(n)=\fpr{\G_{g};P_{1};[z]_{n};E},
\end{equation*}
as well as the corresponding extended moduli spaces $\cM_{g}^{*}(n,m)$ and $\cM_{g}^{*}(n)$.

\subsection*{Construction of algebraic orbits}

A general approach for constructing algebraic orbits in the higher 
genus case is the following. First, choose a leaf $\cV(n,m)$ in 
$\cM_{g}^{*}(n,m)$. The precise definition of such leaf depends on 
the choice of the normalization. Then introduce special coordinates 
$T_{A}$, called \emph{Whitham times}, on $\cV(n,m)$, and define the 
prepotential $S$ of the leaf by the formula 
\begin{equation*}
    dS = Q dE.
\end{equation*}
\emph{Whitham differentials} $d\Om_{A}$ are then obtained from $dS$ by
\begin{equation*}
    \pd_{T_{A}}dS(E|T) = d\Om_{A}(E|T),
\end{equation*}
where we use a connection on $\cN_{g}^{1}$ given by choosing 
$E=\text{const}$ to be horizontal sections. Note that 
$dS=\sum_{A}T_{A}d\Om_{A}$.
The leaf $\cV(n,m)$ is then mapped to the corresponding algebraic orbit
$\cO(n)$ by the following sequence of maps:
\begin{equation}\begin{CD}
\cV(n,m) @>>> \cV(n)  @>>> \cO(n)\\
@. \cup @. \cup\\
@. \cV^{\circ}(n) @. \cO^{\circ}(n)\\
\fpr{\G_{g};P_{1};[z]_{n};E,Q} @>>> \fpr{\G_{g};P_{1};[z]_{n};E} @>>> 
\fpr{\G_{g};P_{1};z_{*}},
\end{CD}\end{equation}
where $\cV^{\circ}(n)$ is a leaf of a subfibration of $\cV(n)$ defined by
fixing the periods of $dE$.  Then, in the coordinates $(E,T)$ on the
``universal curve'' $\cN_{g^{1}}$ over $\cO^{\circ}(n)$, the
differential $dS(E|T)=Q(E|T)dE$ satisfies the equations
$\pd_{A}dS(E|T)=d\Om_{A}(E|T)$. Therefore, $dS$ is a prepotential, $
\pd_{x}\Om_{A}(E|T)=\pd_{T_{A}}p(E|T)$, and we obtain
solutions to Whitham equations.  Note that Whitham coordinates on
$\cV(n,m)$ go to Whitham flows on the algebraic orbit.

For each Whitham derivative $\pd_{T_{A}}$ we define a dual 
``integral'' operator $\oint_{T_{A}}$ in such a way that
\begin{equation}
    \oint_{T_{A}}\pd_{T_{B}}dS=\oint_{T_{A}}d\Om_{B}
    =\oint_{T_{B}}d\Om_{A}=\oint_{T_{B}}\pd_{T_{A}}dS
\qquad \forall T_A,T_B.\label{eq:dualTA}
\end{equation}
The exact definition of such operators $\oint_{T_{A}}$ can be
rather non-trivial and usually includes integration with some weight 
over a cycle and, maybe, some correction terms.
The above identities can be thought of as a generalization of Riemann 
bilinear identities for Whitham differentials, and they are proved 
along the same lines. The $\tau$-function of an algebraic orbit is 
then defined by $\tau(T)=e^{F(T)}$, and
\begin{equation}
    F(T)=\frac12\sum_{T_{A}}T_{A}\oint_{T_{A}}dS.
    \label{F(T):defn}
\end{equation}
It encodes all information about the differentials $d\Om_{A}$ and the 
curve $\G_{g}$, and its first and second derivatives are given by
\begin{align*}
    \pd_{T_{A}} F(T) &= 
    \frac12\oint_{T_{A}}dS+\frac12\sum_{T_{B}}T_{B}\oint_{T_{B}}d\Om_{A}\\
     &=\frac12\oint_{T_{A}}dS+\frac12\sum_{T_{B}}T_{B}\oint_{T_{A}}d\Om_{B}=\oint_{T_{A}}dS,\\
     \pd_{T_{A}T_{B}}F(T)&=\oint_{T_{A}}d\Om_{B}=\oint_{T_{B}}d\Om_{A}.
\end{align*}
Note that the equality of mixed partials of $F(T)$ corresponds to
equation $(\ref{eq:dualTA})$.  Third derivatives of $F(T)$ are then
given by the residue type formula.

\subsection*{Complex-normalized case}

In this case, one chooses
\begin{align*}
t_k&=-\frac1k\res_{P_1}z^kQ\,dE,\qquad  k=1,\dots, n+m,  \\
t_{h,k}&=\iam{k}QdE, \quad t_{E,a_k}=\oint_{a_k}dE,\quad t_{Q,a_k}=\oint_{a_k}dQ, \quad
t_{E,b_k}=\oint_{b_k}dE,\quad  t_{Q,b_k}=\oint_{b_k}dQ, 
\end{align*}
as coordinates on $\cM_{g}(n,m)$. The notation $a_{k}^{-}$ indicates 
that the integral has to be taken over the right side of the cycle.
A \emph{complex} $(3g+n+m)$-dimensional  
$\cV_{\C}(n,m)\subset\cM_{g}^{*}(n,m)$ is defined by imposing the $a$-normalization
condition on the Abelian differentials $dE$ and $dQ$,
\begin{equation*}
\oint_{a_{k}}dE=\oint_{a_{k}}dQ=0\qquad \forall a_{k}\in\cB.\label{eq:c-leaf}
\end{equation*}
Note that the projection of $\cV_{\C}(n,m)$ to $\cM_{g}(n,m)$ is not well-defined.
Whitham coordinates on the leaf are then given by
\begin{alignat*}{2}
T_{k}&=-\frac1k\res_{P_{1}}z^{k}dS,\qquad & \qquad k&=1,\dots,n+m,\\
T_{h,k}&=\oint_{a_{k}}dS,\qquad
T_{E,k}=-\oint_{b_{k}}dQ,\qquad 
T_{Q,k}=\oint_{b_{k}}dE, \qquad & \qquad k&=1,\dots,g
\end{alignat*}

Corresponding Whitham differentials are divided into the following 
four groups.
\begin{itemize}
\item $\pd_{T_{k}}dS(E|T)=d\Om_{k}(E|T)$ are the usual 
complex-normalized Whitham differentials
with prescribed behavior near the puncture,
\begin{equation*}
d\Om_{k}=d\big(\kk^{k}+O(\kk^{-1})\big)\text{ near $P_{1}$}, 
\qquad\qquad \oint_{a_{l}}d\Om_{k}=0.
\end{equation*}
\item $\pd_{T_{h,k}}dS(E|T)=d\Om_{h,k}(E|T)$ are holomorphic
differentials that are dual to the canonical basis of cycles, 
\begin{equation*}
\oint_{a_{i}}d\Om_{h,k}=\dl_{ik}.
\end{equation*}
\item $\pd_{T_{E,k}}dS(E|T)=d\Om_{E,k}(E|T)$ are $a$-normalized and holomorphic everywhere on $\G_{g}$ 
except for the $a$-cycles, where they have jumps,
\begin{equation*}
d\Om_{E,k}(\palpm)=\dl_{kl}dE(\pal),\qquad \oint_{a_{i}}d\Om_{E,k}=0.
\end{equation*}
\item $\pd_{T_{Q,k}}dS(E|T)=d\Om_{Q,k}(E|T)$ are $a$-normalized and holomorphic everywhere on $\G_{g}$ 
except for the $a$-cycles, where they have jumps,
\begin{equation*}
d\Om_{Q,k}(\palpm)=\dl_{kl}dQ(\pal),\qquad \oint_{a_{i}}d\Om_{Q,k}=0.
\end{equation*}
\end{itemize}

The duals $\oint_{T_{A}}$ are then given by
\begin{equation*}
    \oint_{T_{i}}d\Om=\ress \kk^{i}d\Om,\quad
    \oint_{T_{h,k}}d\Om=\frac{-1}{2\pi\I}\Big[\ibb{k}{d\Om}\Big],\quad
    \oint_{T_{E,k}}d\Om=\frac{-1}{2\pi\I}\Big[\iap{k}{Ed\Om}\Big],
\end{equation*}
where the notation $\ibb{k}{d\Om}$ indicates that a certain correction
terms have to be added to make the integral independent on small cycle
deformations.  In particular, 
\begin{equation*}
 \ibb{k}{dS}=\ib{k}{dS}+T_{E,k}E(a_{k}^{+}\cap b_{k}).
\end{equation*}
This remark becomes very important in the real-normalized case and is
discussed at length in the next section.

The third derivatives of $F(T)$ are given by the following theorem
(\cite{Kr-tauWh}).  
\begin{theorem*}    
    \begin{equation*}
        \pd_{T_{A} T_{B}T_{C}}F(T)=
        \sum_{q_s | dE(q_s)=0} \res_{q_s}{\frac{d\Om_A d\Om_B d\Om_C}{dE dQ}}.
    \end{equation*}
\end{theorem*}

\begin{rem}
 The residue type formula for the third derivatives of 
    $F(T)$ implies that if we consider the reduced hierarchy with 
    $dQ=dp$, then $F(T)$ is a solution for the WDVV equation.
\end{rem}

\subsection*{Real-normalized Whitham hierarchies}\label{ss:wh}
\paragraph*{Real leaf and Whitham times}
We begin the study of the real-normalized case by introducing the 
following real-analytic coordinate system on $\cM_{g}(n,m)$:
\begin{equation*}
\tr_k=\re{\frac{-1}k\ress z^k QdE}, \qquad  
\ti_k=\im{\frac{-1}k\ress z^k QdE}, \qquad                              
k=1,\dots,n+m,
\end{equation*}
and
\begin{align*}
\tr_{E,a_{k}}&=\re{\ia{k}{dE}},  &
\ti_{E,a_{k}}&=\im{\ia{k}{dE}},  & 
\tr_{Q,a_{k}}&=\re{\ia{k}{dQ}},  &
\ti_{Q,a_{k}}&=\im{\ia{k}{dQ}},  \\
\tr_{E,b_{k}}&=\re{\ib{k}{dE}},   &
\ti_{E,b_{k}}&=\im{\ib{k}{dE}},   &
\tr_{Q,b_{k}}&=\re{\ib{k}{dQ}},   &
\ti_{Q,b_{k}}&=\im{\ib{k}{dQ}}, \\
\ti_{h,a_{k}}&=\im{\oint_{a^{-}_{k}}QdE}, &
\ti_{h,b_{k}}&=\im{\oint_{b^{+}_{k}}QdE}, &
k&=1,\dots,g.
\end{align*}
Note that all coordinates except $\ti_{h,a_{k}}$, $\ti_{h,b_{k}}$ are 
just the real and imaginary parts of the corresponding complex 
analytic coordinates. A \emph{real leaf} 
$\cV_{\R}(n,m)\subset \cM_{g}(n,m)$ is defined to be a zero set of the 
coordinates
$\tr_{E,a_{k}},\tr_{E,b_{k}},\tr_{Q,a_{k}},\tr_{Q,a_{k}}$.
Alternatively, $\cV_{\R}(n,m)$ can be defined by the following $4g$ basis-independent equations,
\begin{equation}
\re{\oint_{c}dE}=0,\qquad \re{\oint_{c}dQ}=0,\qquad \forall c\in H_{1}(\G_g,\Z)\label{eq:r-leaf}.
\end{equation}
Therefore, $\cV_{\R}(n,m)$ is well-defined as a real-analytic submanifold of 
both $\cM^{*}_{g}(n,m)$ and $\cM_{g}(n,m)$. 

Before introducing Whitham coordinates on $\cV_{\R}(n,m)$, we have to 
make the following important observation. In what follows we work 
with multivalued differentials that usually have the form
$fd\w$, where $f$ is an Abelian integral and the differential $d\w$ can again be multivalued.
These differentials are well-defined only if we cut the surface $\G_{g}$. Thus, we
always assume that some canonical basis $\cB$ of $\G_{g}$ is chosen, and we cut our surface 
along the representatives $a_{k}$, $b_{k}$ of cycles in $\cB$. Sometimes these cuts are not enough
and we need one extra cut $\g$. Let us introduce the following 
notation.
\begin{notation*}
    By $\Ghat$ we denote the normal polygon obtained from  $\G_g$ by
cutting along the $a$ and $b$ cycles. For each cycle $a_{l}$ we denote
by $a^{+}_{l}$ its \emph{left} and by $a^{-}_{l}$ its \emph{right} sides.
Same notation applies to $b$-cycles. The point of intersection of $a^{-}_{l}$ and
$b^{+}_{l}$ cycles is denoted by $\F_{l}$ and the point of intersection of
$a^{+}_{1}$ and $b^{-}_{g}$ cycles is denoted by $\F_{0}$. 
By $\Gcut$ we denote a Riemann surface obtained from $\G_{g}$ by making a cut
$\g$ from $\F_{0}$ to $P_{1}$, and by $\GHcut$ we denote
its normal polygon. Note that topologically $\Gcut\simeq \G_{g}-P_{1}$. 
\end{notation*}

The multivalued differentials that we consider are single valued on either $\Ghat$ or $\GHcut$ and, 
as differentials on $\G_{g}$ or $\Gcut$, they can have jumps across the $a$ and $b$-cycles,
and a cut $\g$. Note that for such differentials, the integral over a cycle depends 
not only on the homology class of the cycle, but also on the side of the cycle and
the actual choice of a representative in the homology class.
Thus, in order to make our construction independent of a choice of such representative in
the homology class, for each differential $fd\w$ we introduce a corresponding cocycle
$[fd\w]$ by the following procedure.  First, we define $[fd\w]$ on basic cycles 
$a_{k}$, $b_{k}$ by adding certain \emph{correction terms} to $\oint fd\w$, and then extend 
it to an arbitrary cycle by linearity. We use the following notation.
\begin{notation*}
For any cycle $c=\a_{1}a_{1}+\cdots \bb_{g}b_{g}$, by
\begin{equation*}
    \oint_{[c]}fd\w=\left[fd\w\right](c)=\a_{1}[fd\w](a_{1})+\cdots+\bb_{g}[fd\w](b_{g})
\end{equation*}
we denote the value of $[fd\w]$ on a cycle $c$. Note that $[fd\w](a_{k})$ and $[fd\w](b_{k})$ still
depend on a side of the cycle. We always choose a right side for $a$-cycles and
left side for $b$-cycles,
\begin{equation*}
    \iaa{k}fd\w:=\iaam{k}fd\w,\qquad \ibb{k}fd\w:=\ibbp{k}fd\w.
\end{equation*}
However, it is convenient to indicate the side explicitly in the intermediate calculations. 
For consistency, we use same notation for
usual Abelian differentials on $\G_{g}$.
In addition, for any differential $df$, we denote by $f_{l}=\int^{\F_{l}}df$
the value of the corresponding Abelian integral $f$ at the point $\F_{l}$ of
intersection of the $a^{-}_{l}$ and $b^{+}_{l}$ cycles.
We also put
\begin{equation*}
f^{\rr}_{l}=\re{f(\F_{l})},\qquad f^{\ii}_{l}=\im{f(\F_{l})}.
\end{equation*}
\end{notation*}
We now define Whitahm times on $\cV_{\R}(n,m)$ by 
\begin{equation*}
\Tr_k=\re{\frac{-1}k\ress z^k QdE},\qquad  
\Ti_k=\im{\frac{-1}k\ress z^k QdE},\qquad  k=1,\dots,n+m
\end{equation*}
and 
\begin{align*}
\Ta{h}{k}&=\im{\iaam{k}QdE},& 
\Tb{h}{k}&=\im{\ibbp{k}QdE}, & 
\Ta{E}{k}&=-\im{\ibb{k} dQ}, \\
\Tb{E}{k}&=\im{\iaa{k} dQ}, &
\Ta{Q}{k}&=\im{\ibb{k} dE}, &
\Tb{Q}{k}&=-\im{\iaa{k} dE}.
\end{align*}
Except for some relabeling, the main change occurs in the definition 
of $\Ta{h}{k}$, $\Tb{h}{k}$:
\begin{equation*}
\Ta{h}{k}=\ti_{h,a_{k}}-\ti_{Q,a_{k}}E^{\rr}_{k},\qquad
\Tb{h}{k}=\ti_{h,b_{k}}-\ti_{Q,b_{k}}E^{\rr}_{k}.
\end{equation*}

\paragraph*{Prepotential}
The prepotential $dS=QdE$ of the real leaf $\cV_{\R}(n,m)$ is 
described by the following proposition.

\begin{proposition}
The prepotential $dS$ is a meromorphic differential on $\G_{g}$,
holomorphic on $\G_{g}-P_{1}$, with pole of order $n+m+1$ at $P_1$, and
with jumps on $a$ and $b$ cycles. Near $P_1$,
\begin{equation*}\label{eq:dS near P1}
dS\sim d\pr{\sum_{j=1}^{m+n} (\Tr_{j}+\I\Ti_{j}) \kk^j + O(z)} + R^S\frac{dz}z,
\end{equation*}
where $z=\kk^{-1}$ is our preferred local coordinate in the 
neighborhood of $P_{1}$, and
\begin{equation}\label{dS:res}
R^S=\ress dS =\frac{1}{2\pi\I}\sum_{l=1}^g\big(\Ta{E}{l}\Tb{Q}{l}-\Tb{E}{l}\Ta{Q}{l}\big).
\end{equation}
The jumps of $dS$ across the $a$ and $b$-cycles come from the jumps of $Q$ 
and are given by
\begin{equation}
dS(\palpm)=\I\Ta{E}{l} dE, \qquad dS(\pblpm)=\I\Tb{E}{l} dE.
\end{equation}
The corresponding cocycle $[dS]$ is well-defined only on 
$\G_{g}-P_{1}$, and is given by
\begin{equation*}
\iaapm{l}{dS}=\iapm{l}{dS}-\I\Tb{E}{l}E_l\qquad 
\ibbpm{l}{dS}=\ibpm{l}{dS}+\I\Ta{E}{l}E_l
\end{equation*}
\end{proposition}
The proof of this proposition is a direct calculation.

Since $dS$ has a residue at $P_{1}$, the corresponding integral 
$S(P)=\int^{P}dS$ is well-defined only on $\GHcut$. 
We choose an additive normalization constant in such a way that the regular part $(S)_{-}$
vanishes at $P_{1}$. We can now consider cocycles corresponding to the differentials 
$EdS$ and $SdE$, which are well-defined on $\G_{g}-P_{1}$.

\begin{proposition}
    The cocycles $[EdS]$ and $[SdE]$ are given by
\begin{align*}
     \iaam{l}{EdS}&=\iam{l}{EdS}+\I\Tb{Q}{l} S_l +\Tb{Q}{l}\Tb{E}{l} E_l 
     -\I\Tb{E}{l}\frac{E_l^2}2,\\
     \ibbp{l}{EdS}&=\ibp{l}{EdS}-\I\Ta{Q}{l} S_l +\Ta{Q}{l}\Ta{E}{l} E_l 
     +\I\Ta{E}{l}\frac{E_l^2}2,\\
     \iaam{l}{SdE}&=\iam{l}{SdE}-\wS^{a^{-}}_{l} E_l+
     \I\Tb{E}{l}\frac{E_l^2}2,\\
     \ibbp{l}{SdE}&=\ibp{l}{SdE}-\wS^{b^{+}}_{l} E_l-
     \I\Ta{E}{l}\frac{E_l^2}2.
\end{align*}
Moreover, the following ``\emph{integration by parts}'' formulas hold:
\begin{align*}
    \re{\iaam{l}{SdE}}&=-\re{\iaam{l}{EdS}+\Ta{h}{l}\Tb{Q}{l}},\\
    \re{\ibbp{l}{SdE}}&=-\re{\ibbp{l}{EdS}-\Tb{h}{l}\Ta{Q}{l}}.
\end{align*}
\end{proposition}

\paragraph*{Whitham differentials}
Similarly to the complex-normalized case, Whitham differentials 
$d\Om_{A}(E|T)=\pd_{T_{A}}dS(E|T)$
 can be divided into four groups:
\begin{itemize}
    \item  $d\Om_{k}^{\rr}$ and $d\Om_{k}^{\ii}$ are real-normalized meromorphic differentials 
    with prescribed singularities at $P_{1}$, i.e., exactly the differentials of the 
    real-normalized Whitham hierarchy,

    \item  $d\Om^{a}_{h,k}$ and $d\Om^{b}_{h,k}$ form a canonical real basis in the space of
    real-normalized holomorphic differentials that is dual to our basis $\cB$,

    \item  $d\Om^{a}_{E,k}$ and $d\Om^{b}_{E,k}$ are meromorphic differentials with a simple
    pole at $P_{1}$ and a $dE$-jump across $a$ and $b$-cycles,

    \item  $d\Om^{a}_{Q,k}$ and $d\Om^{b}_{Q,k}$ are meromorphic differentials with a simple
    pole at $P_{1}$ and a $dQ$-jump across $a$ and $b$-cycles.
    
\end{itemize}

In the sequel we restrict to leaves $\cV_{\R}^{\circ}(n,m)$ of a foliation defined by the level
sets of $\Ta{Q}{k}$, $\Tb{Q}{k}$. Then all the differentials but $d\Om^{a}_{Q,k}$ and 
$d\Om^{b}_{Q,k}$ generate flows preserving the foliation. For this reason, we do not describe
differentials $d\Om^{a}_{Q,k}$,  $d\Om^{b}_{Q,k}$ in detail.
We also need to consider corresponding Abelian differentials, which
we always normalize by $(\Om_A)_(P_1)=0$.

\begin{proposition} For any Whitham time $T_{A}$, the differential
    $d\Om_{A}(E|T)=\pd_{T_{A}}dS(E|T)$  is holomorphic on $\G_{g}-P_{1}$.
\end{proposition}
\begin{proof}
The only possible extra poles can appear when $E(P)$ does not define
a local coordinate, i.e., at the points $q_{s}$ such that $dE(q_{s})=0$. Assuming for simplicity
that $dE$ has a simple pole at $q_{s}$, let $E_{s}(T)=E(q_{s}(T)|T)$, and
choose $\xi(E|T)=\sqrt{E-E_{s}(T)}$ to be a local coordinate near $q_{s}$. Then  
we have
\begin{align*}
    \pd_{T_{A}}dS(E|T)&=\pd_{T_{A}}(Q(\xi(E|T)|T))d(\xi^{2}+E_{s})\\
    &=\frac{dQ}{d\xi}\frac{-\pd_{T_{A}}(E_{s})}{\sqrt{E-E_{s}}}2\xi d\xi
    =-(\pd_{T_{A}}{E_{s}})dQ(E|T),
\end{align*}
which is holomorphic at $q_{s}$. 
\end{proof}

Main properties of the Whitham differentials are summarized in the 
following proposition.
\begin{bproposition}
   \begin{itemize}
       \item    Differentials 
       $d\Om^{\rr}_{k}(E|T)$ and 
       $d\Om^{\ii}_{k}(E|T)$ are \emph{real-normalized},
       \begin{equation*}
           \im{\oint_{c}d\oR{k}}=\im{\oint_{c}d\oI{k}}=0 \qquad \forall c\in H_{1}(\G_{g},\Z),
       \end{equation*}
       meromorphic on $\G_{g}$, with a single pole at $P_{1}$,  
       where
       \begin{equation*}
           d\oR{k}\sim d(\kk^{k}+O(1)),\qquad  
           d\oI{k}\sim d(\I\kk^{k}+O(1))\qquad \ress \oR{k}=\ress \oI{k}=0.
       \end{equation*}
       
       \item  Differentials 
        $d\oah{k}(E|T)$ and 
        $d\obh{k}(E|T)$ 
        are holomorphic on $\G_{g}$ and
        form a canonical basis in the
        space of real-normalized holomorphic differentials dual to our homology basis $\cB$, i.e., 
    \begin{alignat*}{2}
    \im{\ia{l}d\oah{k}}&=\dl_{kl},\qquad&  \im{\ib{l}d\oah{k}}&=0,\\ 
    \im{\ia{l}d\obh{k}}&=0, &       \im{\ib{l}d\oah{k}}&=\dl_{kl}
     \end{alignat*}
     
     \item Differentials 
     $d\oaE{k}(E|T)$ and $d\obE{k}(E|T)$
     are meromorphic on $\G_{g}$,
     holomorphic on $\G_{g}-P_{1}$, have a simple pole at $P_{1}$ 
        which is balanced by a single jump across one of the cycles:
        \begin{alignat*}{2}
            \ress d\oaE{k}&=\frac1{2\pi\I}\Tb{Q}{k}=\frac1{2\pi}\oint_{a_{k}}dE, & \qquad  
            d\oaE{k}(\palpm)&=\I\dl_{kl}dE(\pal),\\
            & & d\oaE{k}(\pblpm)&=0,\\
            \ress d\obE{k}&=\frac{-1}{2\pi\I}\Ta{Q}{k}=\frac1{2\pi}\oint_{b_{k}}dE,  & 
            d\obE{k}(\palpm)&=0,\\
            \phantom{a}& & d\obE{k}(\pblpm)&=\I\dl_{kl}dE(\pbl),
        \end{alignat*}
        differentials themselves  are \emph{not} real-normalized,
        but the corresponding cocycles are,
    \begin{equation*}
        \im{\oint_{[c]}d\oaE{k}}=\im{\oint_{[c]}d\obE{k}}=0.
     \end{equation*}
   \end{itemize}
\end{bproposition}


\paragraph*{Duality and Riemann Bilinear Relations}
In the real-normalized case, the dual ``integral'' operators
$\oint_{T_{A}}$ are given by the following formulas:
\begin{align*}
  \oint_{\Tr_{i}}d\Om &=\re{\ress\kk^{i}d\Om}, &
  \oint_{\Ti_{i}}d\Om&=\re{\ress\I\kk^{i}d\Om}, \\
    \oint_{\Ta hk}d\Om &=\re{\frac{-1}{2\pi}\ibb{k^+}d\Om}, & 
    \oint_{\Ta hk}d\Om &=\re{\frac1{2\pi}\iaa{k^-}d\Om}, \\
    \oint_{\Ta Ek} d\Om
    &=\re{\frac{-1}{2\pi}\left(\iaa{k^-}Ed\Om+\Corr^{a}_{k}(d\Om)\right)}, \\ 
    \oint_{\Ta Ek} d\Om
    &=\re{\frac{-1}{2\pi}\left(\ibb{k^+}Ed\Om+\Corr^{b}_{k}(d\Om)\right)},
\end{align*}
where the correction terms $\Corr(d\Om)$ are given by
\begin{align*}
    \Corr^{a}_{k}(dS)&=\Ta hk\Tb Qk + \Tb hk \Tb Qk, \qquad
    \Corr^{b}_{k}(dS)=-\Ta hk\Tb Qk + \Tb hk \Tb Qk,\\
    \Corr(d\Om_{A})&=\pd_{T_{A}}\Corr(dS).
\end{align*}
We now have to establish relations
$(\ref{eq:dualTA})$.  
\begin{proposition} Whitham differentials satisfy the identity
    \begin{equation}
        \oint_{T_{A}}d\Om_{B}=\oint_{T_{B}}d\Om_{A}\label{rbil}
    \end{equation}
\end{proposition}
\begin{proof}
This proposition is proved by a direct calculation.
We illustrate it by considering the following cases.

First, let $T_{A}=\Tr_{i}$ and $T_{B}=\Tr_{j}$.  Then we have to prove
the following identity:
\begin{equation*}
    \oint_{\Tr_{i}}d\oR{j}=\re{\ress\kk^{i}d\oR{j}} = \re{\ress\kk^{j}
    d\oR{i}}=\oint_{\Tr_{j}}d\oR{i}.
\end{equation*}
But
\begin{align*}
    \ress \kk^{i} d\oR{j} &=-\ress \oR{j}d\kk^{i} 
    =-\ress\oR{j}d\oR{i} -\ress \oR{j}d(\kk^{i})_{-} \\ 
    &=-\ress\oR{j}d\oR{i} -\ress \kk^{j}d(\kk^{i})_{-} 
    =-\ress\oR{j}d\oR{i} +\ress(\kk^{j}d\oR{i}).
\end{align*}
 Taking real parts and observing that
 \begin{equation*}
     \re{\ress\oR{j}d\oR{i}}=\re{\frac{1}{2\pi\I}\oint_{\pd \Ghat}\oR{j}d\oR{i}}=0,
 \end{equation*}
  we obtain the desired identity.
  
Let $T_{A}=\Ta hk$, $T_{B}=\Ta hl$.  Then
\begin{equation*}
    \oint_{\Ta hk}d\oah{l}=\re{\ibbp{k}d\oah{l}}=\re{\ibbp{l} d\oah{k}}=
   \oint_{\Ta hl}d\oah{k}
\end{equation*}
is just the usual Riemann bilinear identity for a canonical basis of
real-normalized holomorphic differentials and it is proved in the regular way:
   \begin{equation*}
       0=\re{\ress\oah{k}d\oah{l}}=\re{\frac{1}{2\pi}\left(-\ib{l}d\oah{k}+\ib{k}d\oah{l}\right)}.
   \end{equation*}

   The most difficult identities to establish correspond to $T_{A}=\Ta
   Ek$, $T_{B}=\Ta Es$ and the like, since $d\oaE{k}$, $d\obE{k}$ have
   simple poles at $P_{1}$.  In this case, let $C_\e$ be a small circle
   around $P_{1}$, $\F_{\e}\in C_\e$, and let $\g_{\e}$ be a cut from
   $\F_{0}$ to $\F_{\e}$. We consider the integral $\oaE{k}d\oaE{s}$ along the contour
\begin{equation}
\sum_{l=1}^g (a_i^+ +b_i^+ -a_i^- -b_i^-) +\g_\e^+ - \g_\e^- + C_\e.
\end{equation}
First, we compute that 
\begin{align*}
&\re{\frac1{2\pi\I}\sum_{l=1}^g\left(\ia{l^+}+\ib{l^+}-\ia{l^-}-\ib{l^-}\right)
\oaE{k}d\oaE{s}}\\
&\qquad\qquad=\re{\frac1{2\pi}\left(\iam{k}Ed\oaE{s}+\iam{s}\oaE{k}dE+\dl_{ks}E^\rr_k\Tb
Qk -\left(\iaam{s}d\oaE{k}\right) E^\rr_s\right)}\\
 &\qquad\qquad=\re{\frac1{2\pi}\left(\iaam{k}Ed\oaE{s}+\iaam{s}\oaE{k}dE
+\Tb Qk((\oaE{s})^\ii_k+\dl_{ks}E^\rr_s)\right)}.
\end{align*} 
Note that
\begin{align*}
    (\oaE{s})^{\ii}_{k}+\dl_{ks}E^{\rr}_{k}&=(\oaE{s})^{\ii}_{0}+
    \im{\int_{\F_{0}}^{\F_{k}}d\oaE{s}}+\dl_{ks}E^{\rr}_{k}\\
    &=(\oaE{s})^{\ii}_{0}=\im{\oaE{s}{(\F_{0}^{-})}}=\im{\oaE{s}{(\F_{0}^{+})}},
\end{align*}
since the only non-trivial imaginary contribution for periods of $d\oaE{s}$ comes from
$b$-periods, and in order for $b^{+}$-contribution not to be canceled by $b^{-}$-contribution,
we need $k=s$, in which case
$
    \im{\int_{\F_{0}}^{\F_{k}}d\oaE{s}}=-\dl_{ks}E^{\rr}_{k}.
$

On the cut $\g$ we have
\begin{equation*}
    \pr{\int_{\g_{\e}^{+}}-\int_{\g_{\e}^{-}}}(\oaE{k}d\oaE{s})=
    \int_{\F_{0}^{+}}^{\F_{\e}^{+}}\Tb{Q}{k}d\oaE{s}=\Tb{Q}{k}(\oaE{s}(\F_{\e}^{+})-\oaE{s}(\F_{0}^{+})).
\end{equation*}

To evaluate $\int_{C_\e}\oaE{k}d\oaE{s}$, we rewrite everything in
terms of a local coordinate $z$ in the neighborhood of $P_1$:
\begin{equation*}
d\oaE{k}=\frac{\Tb{Q}{k}}{2\pi\I} \frac{dz}{z}+(\a+O(z))dz, \qquad
\oaE{k}=\frac{\Tb{Q}{k}}{2\pi\I}\Log(z)+O(z).
\end{equation*}
Then
\begin{align*}
\int_{\F_{\e}^{+}}^{\F_{\e}^{-}}\oaE{k}d\oaE{s}&=
\int_{\F_{\e}^{+}}^{\F_{\e}^{-}}\frac{\Tb{Q}{k}\Tb{Q}{s}}{-4\pi^{2}}\frac{\Log(z)}{z}dz+O(\e)\\
&=\frac{\Tb{Q}{k}\Tb{Q}{s}}{8\pi^{2}}\pr{(\Log(z))^{2}\Big|_{\F_{\e}^{-}}^{\F_{\e}^{+}}}+O(\e)\\
&=\frac{\Tb{Q}{k}\Tb{Q}{s}}{8\pi^{2}}(4\pi\I\Log(z(\F_{\e}^{+}))+4\pi^{2})+O(\e)\\
&=-\Tb{Q}{k}(\oaE{s}(\F_{\e}^{+}))+\frac{\Tb{Q}{k}\Tb{Q}{s}}{2}+O(\e).
\end{align*}

Collecting all of the above together, we obtain
\begin{align*}
0&=\re{\frac1{2\pi\I}\left(
\sum_{l=1}^g\left(\ia{l^+}+\ib{l^+}-\ia{l^-}-\ib{l^-}\right)
+\int_{\g_\e^+} - \int_{\g_\e^-}+\int_{C_\e} \right) 
\oaE{k}d\oaE{s}}\\ 
 &=\re{\frac1{2\pi}\Big(\iaam{k}Ed\oaE{s}+\iaam{s}\oaE{k}dE+O(\e)\Big)}\\
  &\rightarrow
  \re{\frac1{2\pi}\left(\iaa{k^-}Ed\oaE{s}+\iaa{s^-}\oaE{k}dE\right)}
  \qquad \text{as }\e\to0,
\end{align*}
which proves our identity.  All other cases are somewhat intermediate
in difficulty to the cases considered above and are proved along the
same lines.
\end{proof}

The third derivatives of $F(T)$ are given by the following
theorem.
\begin{theorem}
    \begin{equation*}
        \pd_{T_A T_B T_C} F(T) =
        \re{\sum_{q_s | dE(q_s)=0} \res_{q_s}{\frac{d\Om_A d\Om_B d\Om_C}{dE dQ}}}    
    \end{equation*}
\end{theorem}
\begin{proof}
    First we prove the following formula.  For any two Whitham differentials
    $d\Om_{B}$, $d\Om_{C}$ we have
\begin{equation}
    \begin{split}
        \re{\frac1{2\pi\I}\oint_{\pd \hat\G}(\pd_{T_A}\Om_B)d\Om_C}&=\re{\ress(\pd_{T_A}\Om_B)d\Om_C}\\
        &\qquad +
\re{\sum_{q_s | dE(q_s)=0}\res_{q_s}{\frac{d\Om_A d\Om_B d\Om_C}{dE dQ}}}.
        \label{res:real}
    \end{split}
    \end{equation}
    To establish this identity, note that
    since $d\Om_{C}$ is holomorphic outside of $P_{1}$, the  right hand side of our equation is a 
    sum of residues at $P_{1}$ and at poles of $\pd_{T_{A}}\Om_{B}$. Using $E$ as a coordinate, we see
    that 
    \begin{equation*}
        \pd_{T_{A}}\Om_{B}(E|T)=\pd_{T_{A}}\pd_{T_{B}}S(E|T),
    \end{equation*}
    is holomorphic. So the only extra poles can appear at $q_{s}$ such that $dE(q_{s})=0$. Assuming that
    $q_{s}$ is a simple zero of $dE$ and using $\xi(P|T)=\sqrt{E(P)-E_{s}(T)}$ 
    as a local coordinate near $q_{s}$, 
    we have
    \begin{align*}
    \Om_B(P|T)&=\Om_B(q_s(T))+\frac{d\Om_B}{d\xi}(q_s(T))\xi(P|T)+O(\xi(P|T)^2)\\
    (\pd_{T_A}\Om_B)(P|T)&=\frac{d\Om_B}{d\xi}(q_s(T))\frac{-\pd_{T_A}E_s(T)}{2\xi(P|T)}
    + O(1)\\
    (\pd_{T_A} Q)(P|T)&=\frac{dQ}{d\xi}(q_s(T))\frac{-\pd_{T_A}E_s(T)}{2\xi(P|T)}+ O(1).
    \end{align*}
    At the same time,
    \begin{equation*}
        (\pd_{T_A} Q)(P|T)=\pd_{T_A}\frac{dS(P|T)}{dE(P)}=\frac{d\Om_A}{dE}(P|T).
    \end{equation*}
    Therefore,
\begin{align*}
    (\pd_{T_A}\Om_B)(P|T)&=\frac{d\Om_B}{d\xi}\frac{d\xi}{dQ}(q_s(T))\left(\frac{d\Om_A}{dE}(P|T)
    + O(1)\right)\\
    &=\frac{d\Om_B}{dQ}(q_s(T))\frac{d\Om_A}{dE}(P|T) + O(1)
\end{align*}
and we have
\begin{align*}
\res_{q_s} (\pd_{T_A} \Om_B ) d\Om_C &= 
\res_{q_s} \frac{d\Om_B}{dQ}(q_s(T))\frac{d\Om_A}{dE}d\Om_C(P|T)\\
&=\res_{q_s}{\frac{d\Om_A d\Om_B d\Om_C}{dE dQ}(P|T)}.
\end{align*}
The rest of the proof is a direct calculation, which we illustrate by three different cases.
 \begin{align*}
            \pd_{T_A}(\pd_{\Tr_i}\pd_{\Tr_j}F(T))&=\re{\ress\kk^id(\pd_{T_A}\oR{j})}
            = \re{\ress \oR{i}d(\pd_{T_A}\oR{j})}\\ 
            &=-\re{\ress\pd_{T_A}\oR{j}d\oR{i} } =  
            \re{\sum_{q_s | dE(q_s)=0}\res_{q_s}{\frac{d\Om_A d\oR{i} d\oR{j}}{dE dQ}}},
\end{align*}
since $\pd_{T_A}\oR{j}$ is holomorphic at $P_1$ and 
$
\re{\frac1{2\pi\I}\oint_{\pd \hat\G_g}(\pd_{T_A}\oR{j})d\oR{i}}=0. 
$
\begin{align*}
    \pd_{T_A}(\pd_{\Ta hl}\pd_{\Ta hk}F(T))&=
    \re{\frac{-1}{2\pi\I}\ibb{k^+}d(\pd_{T_A}\oah{l})}
    =\re{\frac1{2\pi\I}\oint_{\pd \hat\G_g}(\pd_{T_A}\oah{l})d\oah{k}} \\
    &=\re{\sum_{q_s | dE(q_s)=0}\res_{q_s}{\frac{d\Om_A d\oah{k} d\oah{l}}{dE dQ}}},
\end{align*}
since
$
    \ress(\pd_{T_A}\oah{l})d\oah{k}=0.
$
\begin{align*}
    \pd_{T_A}(\pd_{\Ta El}\pd_{\Ta Ek}F(T))&=
    \re{\frac{-1}{2\pi}\iaa{k^-}Ed(\pd_{T_A}\oaE{k})} 
    = \re{\frac1{2\pi}\iaa{l^-}(\pd_{T_A}\oaE{k})dE}\\
    &=\re{\frac1{2\pi\I}\oint_{\pd \hat\G_g}(\pd_{T_A}\oaE{k})d\oaE{l}}\\
    &=\re{\sum_{q_s | dE(q_s)=0}\res_{q_s}{\frac{d\Om_A d\oaE{k} d\oaE{l}}{dE dQ}}},
\end{align*}
since
$
    \ress(\pd_{T_A}\oaE{l})d\oaE{k}=0
$
and $(\pd_{T_A}\oaE{k})$ is well-defined on $\hat\G_g$.
\end{proof}
\paragraph*{Reduction to Frobenius Structures and Solutions to the WDVV equations}
In the real-normalized case we choose the marked variable to be $\Tr_1$. 
It correspond to the direction of unity,
\begin{equation}
\pd_{\Tr_1}=e\sim d\oR{1}.
\end{equation}
In addition, we choose 
\begin{equation}
    dQ = \I d\oR{1},
\end{equation}
and {\it truncate} the hierarchy by considering 
$\Tr_{i},\Ti_{i}\/$ only for $i\leq n$.  Then we have the following theorem.

\begin{theorem}
Given the above assumptions,
\begin{equation}
    \pd_{\Tr_1 T_A T_B} F(T)=\cases
    \frac{i j}n,\qquad & d\Om_A=d\oR{i},d\Om_B=d\oI{j},\text{ and } i+j=n\\
    \frac{i j}n,\qquad & d\Om_A=d\oI{i},d\Om_B=d\oR{j},\text{ and } i+j=n\\
    \frac1{2\pi},\qquad & d\Om_A=d\oaE{k},d\Om_B=d\oah{k}\\
    \frac1{2\pi},\qquad & d\Om_A=d\obE{k},d\Om_B=d\oah{k}\\
    0           \qquad & \text{ otherwise.}      
    \endcases
\end{equation}
Therefore, the coordinates are flat and $F(T)$ is a solution to the WDVV equation.
\end{theorem}

\begin{proof}
Let us rewrite the formula for the third derivative with $ dQ = \I d\oR{1} $:
\begin{align*}
    \pd_{\Tr_1 T_A T_B} F(T)&=
    \re{\sum_{q_s | dE(q_s)=0} \res_{q_s}{\frac{d\Om_A d\Om_B}{\I dE}}}\\
    &=\re{\frac{1}{2\pi \I }\oint_{\pd \hat\G_g 
    }\frac{d\Om_{A}d\Om_{B}}{\I dE}-\ress\frac{d\Om_{A}d\Om_{B}}{\I dE}}.
\end{align*}

First let us see what is happening near the puncture $P_{1}$. 
Since $dE\sim d(z^{-n}+O(1))$ near $P_{1}$, $\frac{dz}{dE}\sim z^{n+1}$ 
has a zero of order $n+1$ at $P_{1}$.  
It is clear that for the truncated hierarchy the only differentials that
can contribute to the $\ress(*)$-term are $d\oR{i}$ and $d\oI{i}$. Assume that near $P_{1}$,
\begin{equation*}
    d\Om_{A}\sim d(\e_{A}z^{-i}+O(1)),\qquad  d\Om_{B}\sim d(\e_{B}z^{-j}+O(1)),\qquad  \e = 1\text{ or }\I.
\end{equation*}
Then
\begin{align*}
    d\Om_{A} d\Om_{B} &= (\e_{A}(-i)z^{-i-1}+O(1))(\e_{B}(-j)z^{-j-1}+O(1))(dz)^{2}\\
    &=\{\e_{a}\e_{B}(ij)z^{-(i+j)-2}+O(z^{-i-1})+O(z^{-j-1})+O(1)\}(dz)^{2}\\
    \re{-\ress\frac{d\Om_{A}d\Om_{B}}{\I dE}}&=\re{\frac{ij}{-n}\I \e_{A}\e_{B}}\dl_{i+j,n}
    =\frac{ij}n\dl_{i+j,n}\dl_{\e_{A}\e_{B},\I}.
\end{align*}

Now let us consider the boundary term 
$\dfrac{1}{2\pi \I }\oint_{\pd \hat\G_g }\frac{d\Om_{A}d\Om_{B}}{\I dE}$. 
For it to be non-zero, at least one of $d\Om_{A}, 
d\Om_{B} $ should
be $d\oaE{k}$ or $d\obE{k}$.  For example, 
let $d\Om_{A}=d\oaE{k}$.  Then, if $d\Om_{B}\neq d\oaE{s},d\obE{s}$,
\begin{align*}
    \frac{d\oaE{k}}{dE}(\palpm)&=\I \dl_{kl}\qquad \frac{d\oaE{k}}{dE}(\pblpm)=0\\
    \re{\frac{1}{2\pi \I }\oint_{\pd \hat\G_g }\frac{d\oaE{k}d\Om_{B}}{\I dE}}&=
    \re{\frac 1{2\pi\I}\ia{k^{+}}d\Om_{B}}=\cases\frac1{2\pi},\quad &d\Om_{B}=d\oah{k}\\
    0\quad&\text{otherwise}\endcases
\end{align*}
and if $d\Om_{B}=d\oaE{s}$ or $d\Om_{B}=d\obE{s}$,
\begin{equation*}
    \re{\frac{1}{2\pi \I }\oint_{\pd \hat\G_g }\frac{d\oaE{k}d\oaE{s}}{\I dE}}=
    \re{\frac{1}{2\pi \I }\oint_{\pd \hat\G_g }\frac{d\oaE{k}d\obE{s}}{\I dE}}=0.
\end{equation*}
All the remaining cases are similar to the ones considered, 
and this concludes the proof of the theorem.
\end{proof}

\bibliographystyle{amsalpha}

\providecommand{\bysame}{\leavevmode\hbox to3em{\hrulefill}\thinspace}

\end{document}